\documentclass[dvips]{article}

\usepackage{icrc2011}

\title{Segue 1: the best dark matter candidate dwarf galaxy surveyed by MAGIC}

\newcommand{\etal}{\MakeLowercase{\textit{et al. }}} 
\shorttitle{J. Aleksi\'c \etal Segue 1: the best dark matter candidate dwarf galaxy surveyed by MAGIC}

\authors{J. Aleksi\'c$^{1}$, M. Doro$^{2}$, S. Lombardi$^{3}$, D. Nieto$^{4}$ on behalf of the MAGIC Collaboration and \\M. Fornasa$^{5}$}
\afiliations{$^1$Institut de F\'isica d'Altes Energ\'ies, Ed. Cn. Campus UAB, E-08193 Bellaterra, Spain\\ 
$^2$Universitat Aut\`onoma de Barcelona, E-08193 Bellaterra, Spain \\
$^3$Universitat  degli Studi di Padova and INFN, I-35131 Padova, Italy \\
$^4$Universidad Complutense de Madrid, E-28040 Madrid, Spain \\
$^5$Instituto de Astrof\'isica de Andaluc\'ia (IAA-CSIC), E-18080 Granada, Spain \\
  }

\email{jelena@ifae.es}

\abstract{Despite the interest in Dark Matter (DM) searches is currently more focused on underground experiments, a signature of DM annihilation/decay in gamma-rays from space would constitute a smoking gun for its identification. In this contribution, we start with a brief review of the efforts of the ground-based MAGIC Cherenkov telescopes system to detect DM signatures from dwarf satellite galaxies orbiting the Milky Way halo. We then present the recent survey of Segue 1, considered by many as possibly the most DM dominated satellite galaxy known in our galaxy. No significant gamma-ray emission was found above the background in around 30 hours of observation. This is the largest survey ever made on a single dwarf by Cherenkov telescopes. We present a novel analysis that fully takes into account the spectral features of the gamma-ray spectrum of specific DM models in a Supersymmetric scenario. We also discuss the prospects of detection after the Fermi observation of similar objects at lower energies. }

\begin{document}
\maketitle

\section{Introduction}


Over the last century, a great amount of observational evidence at different scales has pointed to the existence of an unknown, invisible mass component that accounts for almost 25\% of Universe's energy budget \cite{r1}. 
Nevertheless, despite strong efforts over the years, Dark Matter is still one of the greatest mysteries of modern Physics. 
No experiment so far has been able to detect DM, directly or indirectly, and shed light on its nature, although recent direct detection experiments might have found a marginal signature of DM \cite{r2,r3}. 
Experimental evidence indicates that the DM particle should be a weakly interacting massive particle (WIMP). 
WIMPs were thermally produced in the early Universe, are stable in time and of non-baryonic nature \cite{r4}. 
Through self-annihilation, WIMPs should produce Standard Model (SM) particles, and some of the by-products, like photons, hadrons and leptons, might be observable from Earth. 
Supersymmetry (SUSY) provides a natural candidate for the DM WIMP - the lightest SUSY particle neutralino.

Given the natural mass range for the SUSY WIMPs (between a few GeV and a few TeV) \cite{r5}, it may be possible to find, via very energetic photons, some signatures of the DM annihilation in the energy range of the Imaging Atmospheric Cherenkov Telescopes (IACTs). 
The typical annihilation gamma-ray spectrum is predicted to be continuous and featureless, due to the gammas mainly being produced from pion decays and final state radiation of charged particles. 
Nevertheless, some distinctive spectral features could be present, like the line emission (from WIMP's annihilation into a pair of gammas or a gamma and a Z-boson), a cut-off or a spectral hardening due to internal bremsstrahlung, all dependent on the mass of the DM particle. 

The gamma-ray flux from DM annihilation that is expected to be measured on Earth is determined by two independent factors: one coming from particle physics and the other related to the astrophysics. 
The particle physics term gives the number of gammas, produced above the given energy threshold for a certain DM model, and it does not depend on the source. 
On the other hand, the astrophysical factor is determined by the DM density profile of a specific source, as well as its distance.
\section{Dwarf Spheroidal Galaxies (dSphs)}

Given that the annihilation rate is proportional to the squared DM density, best places to look for WIMPs are those regions with large concentration of DM.

Dwarf Spheroidals, the satellite galaxies of Milky Way, have been identified as one of the most suitable candidates for DM searches: their ratio between the mass inferred from gravitational effects and the mass inferred from luminosity ($M/L$) can reach values from 100 up to few 1000 $M_\odot/L_\odot$; they are relatively close and they are expected to be free of astrophysical objects whose gamma-ray emission might 'hide' the DM signal \cite{r6}\cite{r7}.  

Until several years ago, only nine dSphs were discovered orbiting the Milky Way. These, so called `classical` dwarfs, were joined by 11 new, ultra-faint objects thanks to the discoveries of the Sloan Digital Sky Survey (SDSS) from 2005 on \cite{r8}. Among these object, several have been identified as excellent candidates for indirect DM searches with MAGIC.

\section{MAGIC Observations}

MAGIC is a ground-based system of two, 17 m diameter IACTs, located at the Observatory Roque de los Muchachos, in the Canary island La Palma (28.8$^{\circ}$ N, 17.8$^{\circ}$ W, 2200 m a.s.l.). MAGIC-I has been operational since 2004, and in 2009 it was joined by MAGIC-II. Together, in stereoscopic mode, they allow for observations of significantly improved sensitivity, lower energy threshold and better energy and angular resolution \cite{r9}.

MAGIC has observed three dSphs so far: Draco \cite{r10}, Willman 1 \cite{r11} and Segue 1 \cite{r12}. All of these observational campaigns were carried out when only the first telescope was operating. Neither case resulted in detection of gamma-ray flux above the background. The flux upper limits (ULs) were calculated, above a certain energy threshold,  for each dwarf and used to probe the parameter space of mSUGRA. mSUGRA is a 5-dimensional subspace of the Minimal Supersymmetric Standard Model (MSSM), defined by universal masses of gauginos ($m_{1/2}$), scalars ($m_0$) and trilinear couplings ($A_0$), as well as by the Higgsino mass term sign ($\mu$) and by the ratio of the vacuum expectation values of two Higgs fields ($\tan\beta$) \cite{r13}. For most mSUGRA models, the lightest supersymmetric particle and DM candidate is the neutralino, a linear combination of the super-partners of the neutral Higgs bosons and gauge bosons.

Findings of these analyses are discussed, for each dSph separately, in the following paragraphs. Segue 1 results are discussed in more detail. 

\subsection{Draco observations}

Draco is one of the nine 'classical' dSphs, located 82 kpc away from us. From the kinematics of over 200 stars associated to this galaxy,  it has been derived that its $M/L$ is around 200 $M_\odot/L_\odot$. 

Draco was observed with MAGIC-I for 7.8 h in 2007. A $2\sigma$ upper flux limit on the steady emission of 1.1$\times10^{-11}$ ph cm$^{-2}$ s$^{-1}$ was found, for energies above 140 GeV, assuming a generic annihilation spectrum without a cut-off and with a spectral index of -1.5. For the purpose of constraining the mSUGRA parameters space, for several benchmark models, we calculated the integral upper limits on expected gamma-ray fluxes for energies above 140 GeV and compared these values to experimental ones. Results showed that ULs obtained from MAGIC Draco observations are $\mathcal{O}(10^{3}$-$10^{9})$ away from constraining the mSUGRA phase-parameter space.

\subsection{Willman 1 observations}

Willman 1 was recently discovered as an ultra-faint dSph, located 38 kpc from us, with $M/L \sim$ 500-700 $M_\odot/L_\odot$. Willman 1 was observed by MAGIC-I in 2008 for 15.5 h. No significant gamma-ray emission was found above 100 GeV,  corresponding to $2\sigma$ integral ULs of order of $10^{-11}$ ph cm$^{-2}$ s$^{-1}$. Assuming the NFW profile of the DM halo, flux was calculated for four different, mSUGRA modified benchmark models \cite{r14}. ULs on the Willman 1 flux measured by MAGIC-I were still $\mathcal{O}(10^3-10^5)$ far from putting constraints on the considered DM models.

\subsection{Segue 1 observations}

Segue 1 is a ultra-faint object discovered in 2007 by the SDSS. Located 23 kpc from the Sun, it was initially considered a globular cluster, but recent results from the kinematics of its 66 stars favour a dSph nature \cite{r16}. In that case, Segue 1 is the most DM dominated object known so far, with $M/L \sim$ 1320-3400 $M_\odot/L_\odot$ \cite{r17}.
 
The Segue 1 observations were carried out with MAGIC-I from late November 2008 until March 2009, for a total of 29.4 hours of good quality data, at low zenith angles, guaranteeing a low energy threshold. This is the longest survey on one dSph conducted by IACTs so far. No significant gamma-ray signal was found above  the background. 

We calculated the integral ULs, for 11 energy bins (above 100 GeV), assuming a power law spectra (with spectral index taking values between -1.0 and -2.4). Obtained values for the ULs are of order $\sim1\times10^{-11}$ ph cm$^{-2}$s$^{-1}$ (above 100 GeV) and $\sim2\times10^{-12}$ ph cm$^{-2}$s$^{-1}$ (above 200 GeV). 

\subsubsection{Constraints}

In order to compare our Segue 1 results with the predictions from mSUGRA, we calculated the expected gamma-ray flux for different models.

The astrophysical term of the flux is model independent, and to estimate its value, we assumed Einasto profile \cite{r18}. Taking into account the source's extended emission, our integration region enclosed $\sim$64\% of the total flux from DM annihilation, so the total and effective astrophysical factor had values of $J(\Delta\Omega)$ = 1.78$\times10^{19}$ GeV$^2$ cm$^{-5}$ sr and $\tilde{J}(\Delta\Omega)$ = 1.14$\times10^{19}$ GeV$^2$ cm$^{-5}$ sr, respectively.

As for the particle physic term, to fully study the mSUGRA phenomenology, we performed a grid scan over the parameter space: for neutralino masses in the detection reach of MAGIC-I, and for both positive and negative sign, we varied the mSUGRA parameters, linearly and in 40 steps over the chosen ranges, to finally obtain over $5\times10^6$ models. Each model was then tested with \texttt{DarkSUSY 5.0.4} for being unphysical and for its compliance with  the SM experimental constraints. Also, for every model, using \texttt{Isasugra 7.78}, we calculated relic density and compared it to the value derived by WMAP \cite{r19} within 3 times its experimental error $\sigma_{WMAP}$. 

 For each DM model in the scan, we computed integral flux UL (above energy threshold $E_0$), using Segue 1 data and the specific gamma-ray spectrum of the particular DM model. Since the spectra for the various models have different shapes and cut-offs, the energy $E_0$ from which the integral flux ($\Phi_{E>E_0}$) was calculated was chosen independently for every tested model. We selected the value of $E_0$ which minimized the {\it expected} flux upper limit, in the case that no excess of events was found.
\begin{figure}[!t]
  \centering
  \includegraphics[width=2.8in]{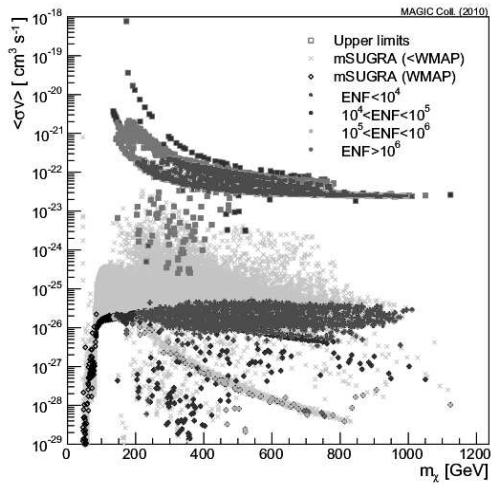}
  \caption{Annihilation cross section ULs from Segue 1 MAGIC data computed for individual points in the scan.}
  \label{fig1}
 \end{figure}

\begin{figure}[!t]
  \centering
  \includegraphics[width=2.8in]{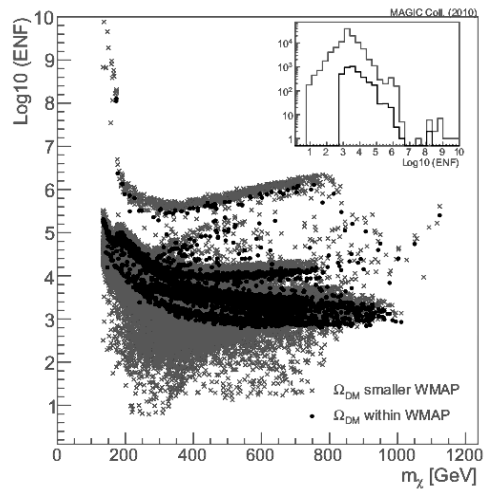}
  \caption{ENFs as a function of the DM mass models in the scan providing a relic density compatible with WMAP scan (circles) or below (crosses).}
  \label{fig2}
 \end{figure}

In order to put constraints on the DM annihilation rate, we translated the calculated ULs (after the $E_0$ optimization) into averaged cross section $\langle \sigma_{ann}v\rangle^{UL}$. We then plotted those values (Figure \ref{fig2}), as a function of $m_{\chi}$, together with the model expectations. Each circle on the plot represents one model, and for each model, there is one UL (square). All the models in the scan for which the relic density is smaller than the value from WMAP are plotted as crosses – for these cases, DM would have to had been produced non-thermally, in order to reach correct relic density. 

Due to the dependence of ULs on the energy spectrum, we found it useful to provide the exclusion plot in terms of enhancement factor (ENF) - a parameter equal to the ratio between the UL on the averaged cross section and the value predicted by mSUGRA. ENF basically indicates how much intrinsic flux boost particular model needs in order to be detectable. Figure \ref{fig2} shows the ENF values, as a function of  $m_{\chi}$, for models whose relic density is compatible with WMAP value or is below it. ENF distribution for these two cases is plotted in the upper panel. It can be seen that models with lowest ENF are the ones with highest annihilation cross section and neutralino mass above 200 GeV, while the majority of points has the ENF greater than $10^4$. For the models compatible with WMAP the lowest ENF is of order $10^3$. It is noticeable that their ENF distribution is quite wide – this is a consequence of the very high ULs relative to the models with small neutralino masses, and in general to the models whose gamma-ray flux above the energy threshold is low. On the other hand, models with relic density below the WMAP value have their annihilation cross sections closer to their upper limits.
 
It is interesting to compare the Segue 1 results of MAGIC-I and Fermi/LAT: due to different energy ranges, Fermi is better suited for low mass neutralinos, while for the heavier DM particles searches with MAGIC are preferred. With the currently available data, Fermi already excludes some mSUGRA models with light neutralino and relic density smaller than the one provided by WMAP \cite{r20}. With 5 years of data it is expected that Fermi could probe some points studied here, and, in case of no detection, exclude most of the large ENF cases. That would leave the space of high-mass neutralinos and relatively small ENF values unconstrained, and that is precisely the region suitable for MAGIC.

For the completeness of this study, it should be pointed out that contribution of monochromatic lines to the annihilation spectrum has been neglected, as expected to be subdominant compared to the continuous emission. In addition, potential presence of substructures in DM halo has not been considered. Their effect on the flux increase is believed to be by a factor few at most. Finally, the particle physics term in the predicted flux does not include the Sommerfeld enhancement. This mechanism is expected to boost the flux value, in some cases, even by a factor of $10^4$. 

\subsubsection{Impact on the PAMELA preferred region}

We have considered the interpretations of the PAMELA positron excess \cite{r21} in the light of DM. According to some DM models, in order to fit PAMELA data, the DM particle should be heavy and annihilate into many leptons, mainly $\mu^+\mu^-$ and $\tau^+\tau^-$. Additional annihilation channel through an intermediate state $\phi$,  that decays into $e^+e^-$, is also considered, although the annihilation cross section required for this process is rather high, $\mathcal{O}(10^{-23}$ - $10^{-24})$ \cite{r22}.
 
\begin{figure}[!t]
  \centering
  \includegraphics[width=2.8in]{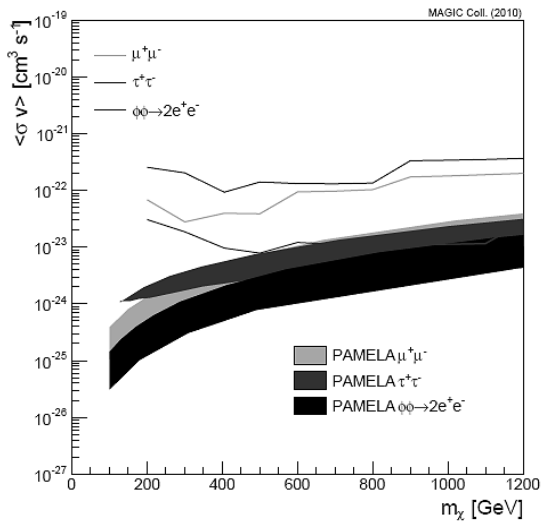}
  \caption{Exclusion lines for a particular annihilation channel and the correspoding regions that provide a good fit to PAMELA data.}
  \label{fig3}
 \end{figure}

We compared our upper limits, obtained for Segue 1 data,  to the regions in the ($m_{\chi}, \langle \sigma_{ann}v\rangle$) plane that provide a good fit for PAMELA data (Figure \ref{fig3}). Annihilation channels $\mu^+\mu^-$, $\tau^+\tau^-$ and $\phi\phi$ (with $m_{\phi}$ = 1 GeV) were considered. These regions have been adapted from \cite{r23}, after rescaling from a local DM density (from 0.3 GeV/cm$^3$ to 0.43 GeV/cm$^3$) \cite{r24}. It can be seen that ENF needed for our ULs to probe the PAMELA-favoured regions is much smaller than in the case of mSUGRA. Furthermore, in the case of $\tau^+\tau^-$ annihilation channel, our ULs are probing the relevant region. However, considering the 2$\sigma$ uncertainty in the astrophysical factor, we do not claim exclusion of the of the PAMELA region with our Segue 1 data. Perhaps in the future, with more  more precise value of $J$, we will be able to place that constraint, at least for massive DM candidates annihilating in $\tau^+\tau^-$.

\section{Conclusions}

The observations of dSphs carried out by MAGIC-I have not resulted in detections. The acquired data were used to calculate the integral upper limits on the annihilation gamma-ray flux, and then to test (different) mSUGRA models. In the most optimistic case, for Segue 1 data, the minimal boost of the DM flux needed for detection is ENF $>$ 600, for energies larger than 100 GeV and for models whose relic density is compatible with one derived by WMAP.  

Segue 1 ULs have also been used to probe DM models which try to explain PAMELA data. Current results are probing the PAMELA preferred region for the case of DM annihilation into $\tau^+\tau^-$, but solid exclusion might be confirmed only if the uncertainties in the astrophysical factor for Segue 1 are eventually reduced.

Although the current observations did not find hints of a signal, and are far from putting the constrains on the mSUGRA parameter space, we expect improvement for observations of dSphs carried out with MAGIC in stereoscopic mode. 

\clearpage


\begin{thebibliography}{}

\bibitem{r1} D. Clowe et. al., ApJ, 2006, {\bf648}: L109-L113
 
\bibitem{r2} R. Bernabei et. al., Eur. phys. J., 2008, {\bf C56}: 333-355

\bibitem{r3} E. Aprile et. al., Phys. Rev. Lett., 2010, {\bf105} (131302)

\bibitem{r4} G. Bertone, D. Hooper, J. Silk, Phys. Rept., 2005, {\bf405}: 279-390
\bibitem{r5} G. Bertone: 2010, Particle Dark Matter, ed., Campridge University Press
\bibitem{r6} J. D. Simon and M. Geha, ApJ, 2007, {\bf670}: 313-331 
\bibitem{r7} L. E. Strigari et. al., Phys. Rev., 2007, {\bf D75} (083526)
\bibitem{r8} B. Yanny et. al., Astron. J., 2007, {\bf137}: 4377-4399
\bibitem{r9} P. Colin et. al., Proceedings of the ICRC, 2009

\bibitem{r10} J. Albert et. al., ApJ, 2008, {\bf679}: 428-431

\bibitem{r11} E. Aliu et. al., ApJ, 2009, {\bf697}: 1299-1304

\bibitem{r12} J. Aleksi\'c et. al., 2011, arXiv: 1103.0407

\bibitem{r13} H. P. Nilles, Phys. Rep., 1984, {\bf110}: 1-162
\bibitem{r14} T. Bringmann, M. Doro and M. Fonasa, JCAP, 2009, {\bf01} (016)


\bibitem{r16} J. D. Simon et. al., ApJ, 2011, {\bf733}: 46 

\bibitem{r17} R. Essig et. al., Phys. Rev., 2010, {\bf D82} (123503)

\bibitem{r18} J. F. Navarro et. al., Mon. Not. Roy. Astron. Soc., 2010, {\bf402}: 21-34

\bibitem{r19} E. Komatsu et. al., ApJ Suppl., 2011, {\bf192}: 18

\bibitem{r20} A. A. Abdo et. al., ApJ, 2010, {\bf712}: 147
 
\bibitem{r21} I. Cholis et. al., JCAP, 2009, {\bf0912} (007)   


\bibitem{r22} N. Arkani-Hamed et. al., Phys. Rev., 2009, {\bf D79 }(015014)

\bibitem{r23} P. Meade et. al., Nucl. Phys., 2010, {\bf B831}: 178-203
 
\bibitem{r24} P. Salucci et. al., A\&A, 2010, {\bf523} (A83)

\end{thebibliography}
\end{document}